\documentclass[
reprint,
onecolumn,
amsmath,amssymb,superscriptaddress,aps,
prl]{revtex4-1}

\usepackage[utf8]{inputenc}
\usepackage{natbib}
\usepackage{graphicx}
\usepackage{floatflt}
\usepackage{hyperref}
\usepackage{url}
\usepackage{siunitx}

\usepackage{xcolor}
\usepackage[normalem]{ulem}

\usepackage{mathtools}

\begin{document}

\title{Number fluctuations induce persistent congestion}

\date{\today}
\author{Verena Krall}
 \email[Electronic Address: ]{verena.krall@tu-dresden.de}
\affiliation{Chair for Network Dynamics, Institute for Theoretical Physics and Center for Advancing Electronics Dresden (cfaed), Technical University of Dresden, 01062 Dresden}

\author{Max F. Burg}
\affiliation{Institute for Theoretical Physics and Centre for Integrative Neuroscience, University of Tübingen, Germany}
\affiliation{Bernstein Center for Computational Neuroscience, Tübingen, Germany}

\author{Malte Schröder}
\affiliation{Chair for Network Dynamics, Institute for Theoretical Physics and Center for Advancing Electronics Dresden (cfaed), Technical University of Dresden, 01062 Dresden}

\author{Marc Timme}
 \email[Electronic Address: ]{marc.timme@tu-dresden.de}
\affiliation{Chair for Network Dynamics, Institute for Theoretical Physics, Center for Advancing Electronics Dresden (cfaed) and Cluster of Excellence Physics of Life, Technical University of Dresden, 01062 Dresden}

\begin{abstract}
The capacity of a street segment quantifies the maximal density of vehicles before congestion arises. Here we show in a simple mathematical model that fluctuations in the instantaneous number of vehicles entering a street segment are sufficient to induce persistent congestion. Congestion emerges even if the average flow is below the segment's capacity where congestion is absent without fluctuations. We explain how this fluctuation-induced congestion emerges due to a self-amplifying reduction of the average vehicle velocities.

\end{abstract}

\maketitle

\section{Research question and hypothesis}

Stationary traffic flows of vehicles on street segments are commonly characterized by the average number of vehicles passing the segment per unit time \cite{manual2000highway}. In reality, vehicles enter and leave a segment at discrete points in time such that the density of vehicles on the segment fluctuates over time. What is the effect of these fluctuations on traffic flow? We hypothesize that stochastic number fluctuations alone may induce congestion.

\section{Methods and Data}
Vehicles travelling at a free-flow velocity $u_f$ pass through a street segment of length $l_0$ in a time $t_0 = l_0 /u_f$. In the presence of other vehicles, this time increases. We model the travel time of a vehicle entering a street segment with a current vehicle density $k$ using Greenshields' model \citep{greenshields1935study, rakha2002comparison}

\begin{align}
    t_{\text{travel}}(k)=\frac{l_0}{u_f}\frac{k_j}{k_j-k} \quad\text{for}\;k < k_j\ , \label{ttravel}
\end{align}

where $k_j$ is the jam density of the street segment: when $k\geq k_j$\,, the travel time diverges as the velocity of all vehicles drops to zero.

A simple macroscopic model of this process describes the change of the average density $\left< k(t) \right>$ of vehicles on the street segment over time as the difference between the incoming flow of vehicles $q_\text{in}$ and the outgoing flow $q_\text{out}(\left< k(t) \right>)=\left<k(t)\right> \, u(\left<k(t)\right>)$,

\begin{align}
    \frac{\mathrm{d} \left< k(t) \right>}{\mathrm{d}t} = \frac{1}{l_0}\left\lbrack q_\text{in}-q_\text{out}(\left< k(t) \right>)\right\rbrack 
    =\frac{1}{l_0} \left\lbrack q_\text{in} - \left(u_f \left< k(t) \right> -\frac{u_f}{k_j}\left< k(t) \right>^2\right)\right\rbrack \ , \label{eq:inout}
\end{align}

where we substituted the average velocity $u(\left<k(t)\right>) = l_0 \lbrack t_\text{travel}(\left<k(t)\right>)\rbrack^{-1}$ using Eq.~\eqref{ttravel}. This model thus neglects stochastic fluctuations of $k(t)$\,.

We contrast this macroscopic model for average vehicle density $\left< k(t) \right>$ with a stochastic model for the instantaneous number density $k(t)$ of vehicles, implemented in an agent-based simulation. Vehicles enter the segment individually according to a Poisson process \cite{TN_libero_mab21251955} with the rate $\nu_\text{in}=q_\text{in}$. Each vehicle has a velocity which is determined by the present vehicle density when it enters the segment and stays constant over time. We neglect here temporary changes in the velocity due to, e.g., random braking  \cite{nagel1992cellular} that would not counter the mechanism described here but rather cause even more frequent traffic breakdowns. We also do not consider a speed-up of vehicles when others leave the segment in front of them.

As a specific setting, we consider a one-lane highway (HW) segment of length $l_{0, \text{HW}}=\SI{1}{km}$\, with free-flow velocity $u_{f,\text{HW}}=\SI{120}{km/h}$\,. We estimate the critical density of vehicles $k_c$ simultaneously on the road above which congestion is likely to occur by considering a critical distance between vehicles of $d_\text{crit}\approx \SI{33.3}{m}$\,, given as the distance driven by a vehicle with velocity $u_{f, \text{HW}}$ in one second. This critical density becomes $k_c = 1/d_\text{crit} = 30 \,\si{veh/km}$ in the deterministic setting. 

The jam density is derived by noting that the outflow is maximal at the critical density, thus

\begin{align}
    \left.\frac{\partial q_\text{out}}{\partial k}\right\vert_{k_c}=u_f-\frac{2 u_f}{k_j}k_c  &\overset{!}{=} 0\\\notag
    \Leftrightarrow k_j & =2 k_c\ .
\end{align}

Thus, the critical flow is

\begin{align}
    q_\text{in}^c=q_\text{out}^c=u_f \left(k_c-\frac{k_c^2}{k_j}\right) = \frac{u_f k_c}{2} \ .
\end{align}

\section{Findings}

\begin{figure}[h!]
    \centering
    \includegraphics[width=\textwidth]{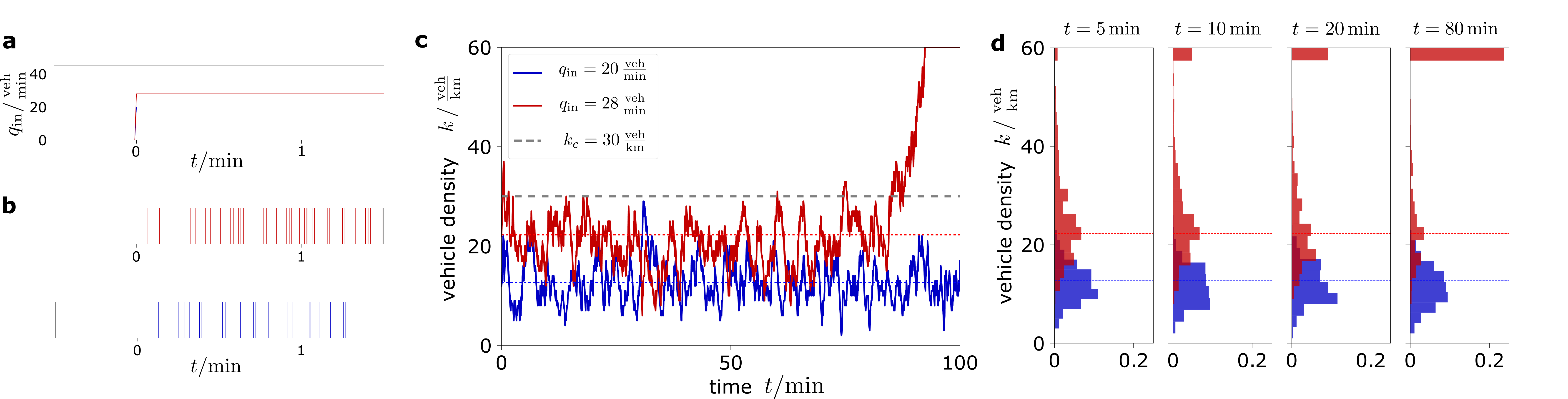}
    \caption{
    The constant in-flow entering the macroscopic model (a) does not capture the stochastic vehicle number fluctuations resulting from a Poisson process (b). However, these fluctuations may cause congestion, even though stable fixed points (indicated by dashed lines) are predicted from the differential equation (c). Considering a sample of 500 random realizations, for $q_\text{in}=28\,\si{veh/min}$ a majority of realizations eventually reaches a congested state (d).}
    \label{fig:vehiclenumbers}
\end{figure}

In the stochastic model, where vehicles enter the street segment in a Poisson process, the time $\Delta t$ between two subsequent vehicles is a random variable drawn from the exponential probability density $\rho(\Delta t) = q_\text{in} \exp{\left(-q_\text{in} \, \Delta t\right)}$ \cite{peebles1987probability}. Hence, the higher $q_\text{in}$\,, the more frequent are small values of $\Delta t$. Several vehicles may enter during a small time interval (Fig. 1b), temporarily increasing the density of vehicles $k(t)$ on the segment above the expected mean value. As a result we observe two different types of dynamics for $k(t)$ during a longer observation interval (Fig. 1c). For low $q_\text{in}$\, the density of vehicles fluctuates near some fixed base level such that traffic flows freely. However, for a sufficiently large in-flow, once $k$ surpasses a threshold value the density of vehicles increases rapidly and the segment congests. When performing repeated simulations we find that such emergent congestions are not negligible outliers but instead are highly probable to occur over time.

\begin{figure}[h]
    \centering
    \includegraphics[width=0.8\textwidth]{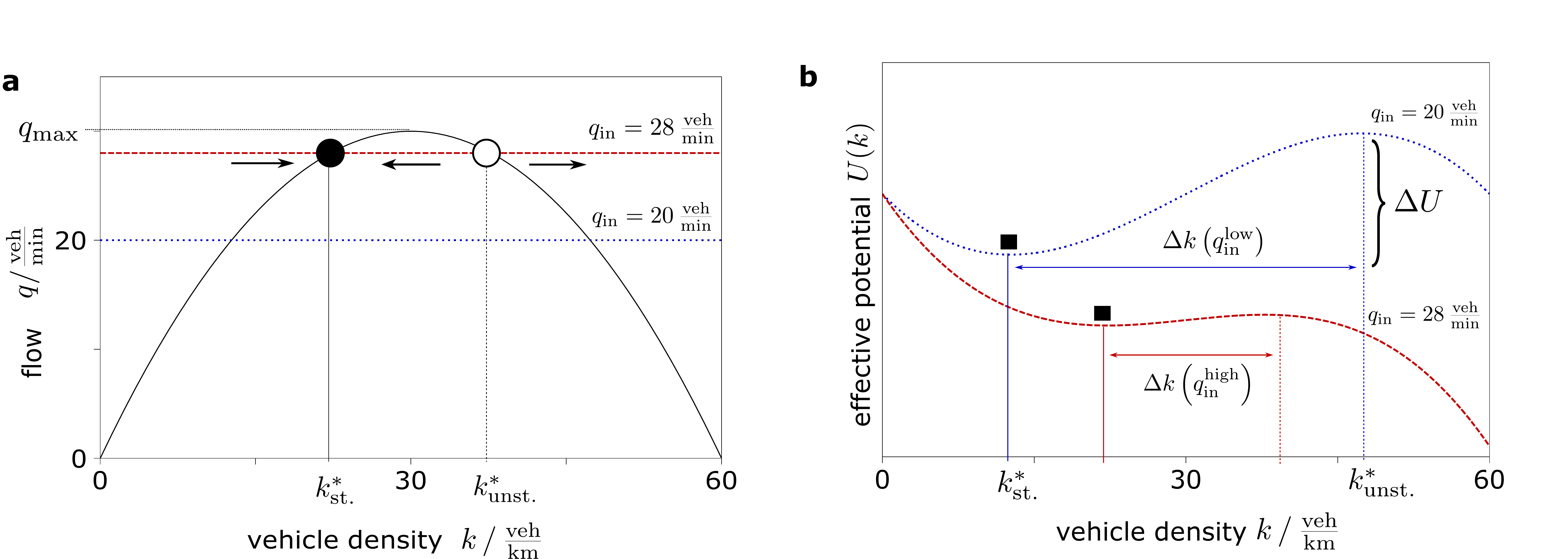}
    \caption{
    The higher the in-flow, the more easily the density of vehicles on the street segment exceeds the threshold to congestion $k_c=k_j/2$. (a) The out-flow-curve defined in equation \eqref{eq:inout} resembles the qualitative form of the well-known fundamental diagram \cite{helbing2001traffic}\,. For each in-rate below $q_\text{max}$\,, a stable and an unstable equilibrium exist (filled and open circle, respectively).
    (b) The effective potential resulting from Eq.~\eqref{eq:inout} resembles a valley-and-peak landscape and the dynamics of the vehicle density is analogous to a block sliding in this potential. On the left side of the valley, representing the stable state, a steep slope pushes the vehicle density away from zero for small $k(t)$. For large $k(t)$ there is a deep canyon, representing a congested state. This landscape changes depending on the flow of vehicles entering the street segment. For small $q_\text{in}$, the stable valley itself is also deep. The block within the valley is prevented from falling into the canyon by a peak of large height $\Delta U$. Thus, the chances that small displacements of the block (small fluctuations of the in-flow of vehicles) push it outside the safe region are very low. With increasing $q_\text{in}$ the distance to the peak and the height $\Delta U$ decrease. Smaller and smaller fluctuations may cause the block to slip over the peak and into the canyon (congested state) more and more easily.}
    \label{fig:explanation}
\end{figure}

How does this spontaneous breakdown of traffic flow emerge? To answer this question, we compare the deterministic model with the stochastic model. In Eq.~\eqref{eq:inout}\,, the out-flow $q_\text{out}(\left<k\right>)$ is a function of the average number density of vehicles with a maximal value $q_\text{max}=q_c$\,, which equals the maximal in-flow. For sufficiently small incoming flows $q_\text{in} \ll q_c$, we find that small temporary changes in the instantaneous number density of entering vehicles are counterbalanced by changes in the instantaneous number of vehicles leaving the segment per time, reflecting a stable equilibrium flow (Fig. 2a) near a fixed point $k_\text{stable}^\ast$\ of the differential equation Eq.~\eqref{eq:inout}. Fluctuations of the number density of vehicles are small and thus, on average, the equilibrium value serves well as a prediction of the traffic state. However, already for $q_\text{in}$ below the maximal in-flow $q_c$ set by the macroscopic model, small changes in $k(t)$ induce congestion: Once a second, unstable equilibrium $k^\ast_\text{unstable}$ is exceeded, the instability pushes $k(t)$ to grow further. The probability of return below $k^\ast_\text{unstable}$ decreases as the density of vehicles increases, thus causing the congestion to be persistent.

Roughly speaking, with an increasing density of vehicles on the segment, their velocities become lower so rapidly that the number of vehicles leaving the segment is persistently lower than the number of vehicles entering during a given time interval, thus causing further velocity reductions in a self-amplifying process. In reality, the system ultimately reaches a fully congested state with vehicles lining up throughout the segment.

We can intuitively understand why the unstable state becomes more important for larger values of $q_\text{in}$: consider the right-hand side of Eq.~\eqref{eq:inout}. As we are interested in the dynamics of vehicle density when perturbed from equilibrium, we picture a sliding block whose position corresponds to the density value $k$ (Fig. 2b). In this image, the differential equation \eqref{eq:inout} would determine the velocity of the block, which equals the negative derivative of an effective potential, a valley-and-peak landscape, with a minimum at $q_\text{stable}^\ast$ and a maximum at $q_\text{unstable}^\ast$ (Fig.~2b).
This potential is defined via
\begin{align}
    \frac{\mathrm{d} U(k)}{\mathrm{d} k} = U'(k):=-q_\text{in}+u_f k \left(1-\frac{k}{k_j}\right)\ . \label{pot}
\end{align}

In our traffic flow scenario, the height difference $\Delta U$ and the distance between the peak and valley reflect the proximity of the two equilibria. The closer the in-flow $q_\text{in}$ is to the maximal flow $q_c$, the closer the two equlibria are and the smaller the vehicle number density fluctuations needed to end up in a congested state. Thus, the randomness of the vehicle-entering process crucially underlies the spontaneous transition from free flow to congestion. 

Moreover, these fluctuations become increasingly more likely: Based on the general theory of stochastic processes \citep{van1992stochastic, TN_libero_mab21251955}, we predict a mean time $T_\text{esc}$ after which a transition to congestion occurs. For a one-lane-highway with $q_c=1800$ vehicles per hour (see Methods), congestion emerges roughly once every 32 hours already if the in-flow is $15\%$ below the maximal in-flow, i.e., $q_\text{in}=1530$ vehicles per hour. If the in-flow is $10\%$ below the maximal value ($q_\text{in}=1620$ vehicles per hour), the transition is expected to occur already after just 2.75 hours.

There exist several studies discussing the instability of traffic flow caused by stochasticity \cite{treiber2013traffic}, for example, due to random route choices \cite{andreotti2015modeling} or the emergence of vehicle clusters due to random distance fluctuations \cite{kuhne2002probabilistic}. 
Typically, explanations for traffic fluctuations are random decelerations due to imperfect human driving behavior \cite{nagel1992cellular}, external factors such as weather conditions, daytime \cite{kim2010likelihood} or the existence of bottlenecks \cite{kerner2019complex}. 
In contrast, our minimal model shows that spontaneous number fluctuations alone already suffice to induce congestion. Although the aforementioned factors that influence traffic have to be considered for (the prediction and prevention of) traffic jams, we thus emphasize that even in the absence of such external influences, congested states may arise due to pure number fluctuation.
As neither the model details for the stochastic process nor those for the travel time are relevant for the fundamental mechanism inducing the presented transition \cite{TN_libero_mab21251955}, we conclude that congestion may emerge spontaneously even for, in general, subcritical traffic flows.
Hence, for analyzing traffic flow and planning infrastructure, one needs to go beyond average quantities and take into account pure number fluctuations.
Future investigations should aim at quantifying the effect of number fluctuations in isolation, by analyzing the emergence of traffic jams in the absence of other external influences on traffic flow. A promising setup may be a highway where traffic consists only of autonomous vehicles, to ensure that imperfect driving behavior does not have to be taken into account in the analysis. Furthermore, a thorough study of the described mechanism in the presence of varying traffic conditions would be particularly valuable for  estimating the relevance of vehicle number fluctuations in real-world settings.

\section{Acknowledgments}
We thank Dirk Witthaut for valuable discussions. The authors thank the German Academic Scholarship Foundation for its role in supporting the initiation of the project. V.K. thanks the Evangelisches Studienwerk. M.F.B. thanks the International Max Planck Research School for Intelligent Systems (IMPRS-IS). 

\section{Data availability}
All material has been made publicly available at GitHub and can be accessed at \url{https://doi.org/10.5281/zenodo.4314973}\,.

\bibliographystyle{plain}
\bibliography{references}

\end{document}